# Modeling of long-term defect evolution in heavy-ion irradiated 3C-SiC: Mechanism for thermal annealing and influences of spatial correlation

Daxi Guo [1], Ignacio Martin-Bragado [2*], Chaohui He [1*], Hang Zang [1], Peng Zhang [1]

[1] *School of Nuclear Science and Technology, Xi'an Jiaotong University, Xi'an, 710049, China*
[2] *IMDEA Materiales, C/ Eric Kandel, 2, Tecnogetafe, 28906 Getafe, Madrid, Spain*

Based on the parameters from published ab-inito theoretical and experimental studies, and combining Molecular Dynamics (MD) and kinetic Monte Carlo (KMC) simulations, a framework of multi-scale modeling is developed to investigate the long-term evolution of displacement damage induced by heavy-ion irradiation in cubic silicon carbide. The isochronal annealing after heavy ion irradiation is simulated, and the annealing behaviors of total interstitials are found consistent with previous experiments. Two annealing stages below 600K and one stage above 900K are identified. The mechanisms for those recovery stages are interpreted by the evolution of defects. The influence of the spatial correlation in primary damage on defect recovery has been studied and found insignificant when the damage dose is high enough, which sheds light on the applicability of approaches with mean-field approximation to the long-term evolution of damage by heavy ions in SiC.

## I. Introduction

The wide band gap and high thermal conductivity make silicon carbide a promising material for microelectronic devices applied in harsh environments.[1,2] However, ion implantation during device fabrication, as well as heavy-ion irradiation in space application will inevitably induce displacement defects. Such defects can undergo migration, recombination and aggregation in the long term, leading to degradation in electronic,[3-5] and thermal[6,7] properties as well as dimensional instability.[8,9] The long-term evolution of displacement damage by heavy ions is therefore fundamental in the fabrication and application of SiC devices.

Thermal annealing is an important process applied to reduce the displacement damage in materials and has been generally studied by isochronal annealing, where distinct recovery stages can be identified. Rutherford backscattering spectroscopy (RBS/C) measurements of heavy-ion-irradiated SiC during isochronal annealing have identified two annealing stages below 600K at low ion fluence and another stage at about 650K at high ion fluence.[10-12] To interpret those annealing stages, theoretical studies have been done on the energetics of point defect and point defects reactions. Using MD simulations, Fei Gao and Weber et al. have studied the recovery of close Frenkel pairs,[13] and the migration of intrinsic point defects of 3C-SiC,[14] and presented an overall mapping of defect recovery in SiC. Bockstedte et al. have done comprehensive studies on the formation and migration of point defects,[15] the recombination of vacancy-interstitial pairs and the aggregation of carbon interstitials[16] of 3C-SiC by ab initio calculations, and proposed a hierarchy of defect recovery.[16] Zheng et al.[17] have recently studied the energy barriers of key point defect reactions based on ab initio molecular dynamics (AIMD) and correlated those processes to the mechanisms of annealing stages. However, the energetics of defects and defect reactions, cannot sufficiently interpret defect recovery, considering that it is only valid for dilute defects without significant clustering[17] and multiple mechanisms can be active simultaneously. Therefore the thermal recovery of radiation damage needs to be modeled kinetically.

The production and evolution of radiation damage involve processes with different temporal and spatial scales, which need to be simulated with different methods. For neutron and heavy ion irradiation, displacement damage is produced in the form of displacement cascades. The displacement cascades evolve and equilibrate over time on the scale of 10ps, and they are confined in a volume of tens of nanometers, which has been simulated by molecular dynamics. Compared to the spherical compact cascades in metals, displacement cascades of SiC are more dispersed and linear,[18] due to high thermal stability and inefficient energy transfer between the atoms of SiC.[18] Cascades by high energy recoils in SiC have been found to split into multiple sub-cascades, with low density cores.[19] As reported in the literature,[20] there is no significant dependence of irradiation temperature on the production of vacancies and antisite defects within the timescale of MD simulation. For long-term evolution, involving i.e. long-range migration of interstitials and vacancies, MD is computationally unsuitable. KMC has been extensively

*Corresponding Author: Email addresses: ignacio.martin@imdea.org. Tel.: +34 91 549 34 22; Fax: +34 91 550 30 47 and hechaohui@mail.xjtu.edu.cn. Tel.: +86 13201420820; Fax: +86 29 82667802

applied to simulate the long-term evolution of radiation damage in metals.[21-27] However, KMC simulation for defect recovery in SiC is limited. Using kinetic lattice Monte Carlo (KLMC) and based on parameters from MD simulation, Rong et al. has studied the defect recovery in 3C-SiC, extended the annealing time of a MD cascade from 10ps to minutes,[27] although only a single cascade was annealed. To mimic the case of experiments, multiple cascades should be included. However, for KMC simulations, the maximum simulation volume size is limited to microns and the maximum damage dose is typically less than 1dpa.[28] With the mean-field approximation, the mean filed rate theory (MFRT) calculation treats the evolution of radiation damage by solving an array of Master Equations describing the recombination and clustering of point defects, the growth and the shrinkage of defect clusters,[28] which is suitable to model long-term evolution of radiation damage with damage dose and simulation timescales as high as those within the lifetime of a nuclear reactor. Using a rate theory model based on the parameters of point defects and point defect reactions from reported ab-initio studies, Swaminathan et al. have studied the amorphization of single crystal[29] and nano-grained[30] 3C-SiC by electron irradiation, and found that the defect migration barriers and the defect recombination barriers are important for the amorphization[29] and for the roles of grain refinement on radiation resistance.[30] However, the MFRT calculations do not involve the spatial correlation, which can be important for displacement cascades induced by heavy ions and neutrons, as close I-V pairs and defect clusters can be formed directly in the cascades. J. Dalla Torre et al.[31] have demonstrated that Cluster Dynamics (CD) based on the mean-field approximation cannot reproduce the correlated recombination in the annealing of electron-irradiated iron. Oritz et al.[23] have simulated the long-term annealing of isolated Frenkel pairs and cascade damage in iron by object kinetic Monte Carlo (OKMC) and MFRT. The results indicate that compared to annealing of isolated defects, new recovery peaks at high temperature appear in the case of cascade damage annealing.

In this study, using the generic kinetic Monte Carlo simulator MMonCa,[32,33] which has been recently extended to simulate binary crystal, and combining the results of displacement cascades from MD simulations, and based on parameters from published theoretical and experimental studies, we have developed a framework of multi-scale modeling. The isochronal annealing of heavy-ion-irradiated 3C-SiC was simulated, and the results were compared to those of experiments. Based on the defect evolution during the isochronal annealing, the mechanisms of defect annealing in SiC were also studied. Finally, we discuss the influence of spatial correlation in primary damage on the defect recovery in SiC, to evaluate the applicability of methods like MFRT with mean-field approximation to the long-term evolution of cascade damage by neutrons and heavy ions.

## II. Calculation approach
### A. Calculation of damage profile

In this study, we simulate $1nm^{-2}$ 550keV $Si^+$ implantation into 6H-SiC with incident angle of 30 degree at 160K and the subsequent isochronal annealing up to 1170K.[10] It should be noted that the MD displacement cascades and OKMC simulations in this study were performed based on the configuration and the defect parameters of 3C-SiC. It has been found that the general disordering behavior and the recovery stages from near room temperature to 600K in 3C-SiC are similar to those of 4H-SiC and 6H-SiC.[34] Therefore, it is assumed that the simulation results of 3C-SiC can be compared to the experiments on 4H-SiC and 6H-SiC. Damage profile by the implantation was calculated by SRIM[35] with the "Detailed Calculation with full Damage Cascade" mode, assuming threshold energies for displacement of Si and C atom to be 35eV and 20eV,[36] and using the NRT calculation approach depicted by Stoller et al.,[37] with the average threshold displacement energy for SiC assumed to be 25eV,[18] respectively. In the present work, we chose a dose of 0.023 dpa, close to the peak damage of the implantation, as shown in Fig. 1.

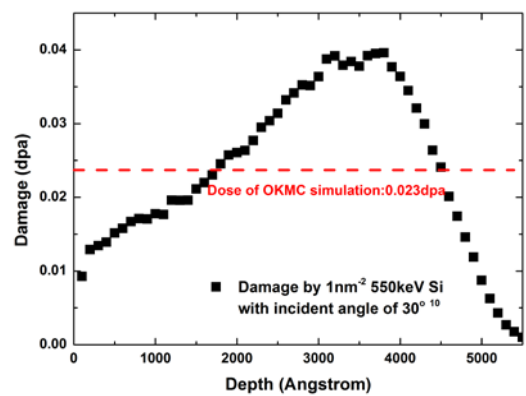

FIG. 1. Damage profile by $1nm^{-2}$ 550keV Si with incident angle of 30°[10] calculated by SRIM.

### B. Calculation of displacement damage by recoils

The formation of displacement cascades in 3C-SiC with the temporal length on the order of 10ps has been modeled by molecular dynamics (MD). In this study, cascades by 10keV Si primary knock-on atoms (PKA) were simulated because higher energy cascades (up to 50keV) involve the formation of subcascades similar in structure to 10keV cascades,[19] with similar clustering fraction and size distribution of defect clusters.[18] In addition, OKMC simulations of iron using Fe cascades by full MD[25] and binary collision approximation (BCA)[26] simulations with PKA energy from 10keV to 200keV (MD)[25] and from 5keV to 100keV (BCA)[26] have revealed that the long-term nano-structural evolutions, i.e. the size distributions of vacancy and interstitial clusters, are almost insensitive to the PKA energy. The results indicate that cascades by low energy PKA can be used as building blocks for higher energy cascades for long-term evolution. In this study, damage by 550keV Si ions in SiC has been linearly decomposed into to a number of cascades by 10keV Si PKA, as described in Section II(C).

Displacement cascades by 10keV Si PKA in 3C-SiC were simulated by molecular dynamics, using LAMMPS.[38] The displacement cascades were simulated at 160K, in a simulation cell of $50a_0 \times 40a_0 \times 40a_0$ (where $a_0$ is the lattice parameter for 3C-SiC) with periodic boundary conditions. The inter-atomic interactions were described by a hybrid Tersoff/ZBL potential.[20,39] 10keV PKAs were introduced on the top-center of the simulation cell in the [4 11 9̄5] direction to avoid channeling.[20,40] After the slowing-down of PKAs within 0.2ps and a 1.8ps short-term annealing, the system was equilibrated for 10ps, as described by Farrell et al.[20] Defects were identified using the Wigner-Seitz cell method, where the occupancies of atoms from output geometry in the Wigner-Seitz cells of reference lattices, are sorted.[20] Lattice sites with zero occupancy are identified as vacancies, cells with occupancy larger than 1 are recognized as interstitial sites, and sites with occupancy equal to 1 but occupied by wrong atoms are considered as antisite defects. With the above procedures, the average number of vacancies is determined to be 120, in agreement with the results of Farrell et al (~120).[20] With the searching radius for clustering set to be the second nearest neighbor (2NN≈3.08Å), the clustering fraction for vacancy is calculated to be 44.7%, which is higher than that in the case of 100K and 500K in Ref. 20 by 30%, probably due to a smaller searching radius (2.2 Å) used in their work. Defect configurations and distributions obtained in the MD displacement cascade simulations were used as inputs for the subsequent OKMC simulations.

**C. OKMC simulation for long-term defect evolution**

The long term evolution of displacement damage was modeled by OKMC[41] using the MMonCa code.[32,33] In the framework of OKMC, each point defect or defect cluster, as well as each impurity atom is defined as an object, which can migrate or interact with each other, or transform into another type.

Since previous ab-initio studies have shown that the as-grown 3C-SiC is generally n-type,[42] we considered the n-type 3C-SiC in this study. Six types of point defects were considered, including vacancies ($V_C$ and $V_{Si}$), interstitials ($C_I$ and $Si_I$), and antisites ($C_{Si}$ and $Si_C$), where $C_{Si}$ refers to a carbon atom situated on a Si lattice site and vice versa for $Si_C$. $C_I$ and $Si_I$ were taken to be the most stable configurations of n-doping 3C-SiC, which are $C_{sp<100>}$ (C-C dumbbell in the <100> direction) and $Si_{sp<110>}$ (Si-Si dumbbell in the <110> direction),[15,16,42,43] respectively.

Formation energies and migration barriers of point defects used in this OKMC simulation are summarized in Table I. Most of these parameters come from a recent ab-initio calculation by D. Shraders et al.,[43] while a few parameters are adopted from other works[44,46] for the following reasons:

1. Ref. 43 has proposed a lower migration barrier of 2.7eV for $V_{Si}$ in n-doping 3C-SiC, which corresponds to a charge state of -2. However, the experimental observed $V_{Si}$ possess a -1 charge state, due to its high-spin configuration (S=3/2), and has been unambiguously identified the T1 center,[47-49] which can be detected in n-type and p-type 3C-SiC irradiated by electrons and protons.[49] Therefore, the migration barrier of 3.2eV for $V_{Si}^{1-}$ reported by M.Bockstedte et al.[15] was adopted.
2. A lower value of about 0.8 eV[43,50] has been reported for the migration barrier of $Si_I$. However, based on the dependence of interstitial loop denuded zone (DZ) width on temperature along grain boundaries in 3C-SiC,[46] the activation energy for migration of Si interstitials has been determined to be about 1.5eV. Therefore, a close value of 1.48eV reported in Ref. 43 was used in this study.
3. For antisite defects, only the migrations by exchange processes are considered, with the migration barriers of 11.6eV and 11.7eV for $C_{Si}$ and $Si_C$, respectively,[44,45] ignoring the vacancy-mediated migration mechanism.[51]

Moreover, the Fermi level can be modified by deep donors or acceptors levels introduced by defects, and a high dose irradiation could transform the n-type material into an intrinsic or p-type one. Should this happen, different sets of formation energies and migration barriers for each point defect should be adopted.[15]

TABLE I. Formation energies ($E_f$), migration barriers ($E_m$), and pre-exponential factors ($D_0$) for point defects

| Defect type | $E_f$ (eV) | $E_m$ (eV) | $D_0$ ($10^{-3}$ $cm^2/s$) |
|---|---|---|---|
| $V_C$ | 4.19[a] | 3.66[a] | 0.743[b] |
| $V_{Si}$ | 4.97[a] | 3.20[a] | 0.743[b] |
| $C_I$ | 6.95[a] | 0.67[a] | 1.230[c] |
| $Si_I$ | 8.75[a] | 1.48[a,d] | 3.300[c] |
| $C_{Si}$ | 4.03[a] | 11.70[d,e,f] | |
| $Si_C$ | 3.56[a] | 11.60[e,f] | |

[a]Ref. 43; [b]Ref. 29; [c]Ref. 14; [d]Ref. 46; [e]Ref.44 ; [f]Ref. 45

In the current OKMC simulations, the following events are considered:

*1. Migration of point defects*

In this study, point defects except antisites can migrate randomly with a specified migration distance λ, while interstitial clusters, vacancy clusters and defect complexes (i.e Frenkel pairs) are assumed to be immobile. Due to the high migration barrier, antisite defects are also assumed to be immobile within the studied temperature range. The migration rate for mobile defects is defined as:

$$\nu_m = \nu_m^0 \exp(-\frac{E_m}{k_B T}) \quad (1)$$

Where $\nu_m^0$ is the attempt frequency, which is given by $(2dD_0/\lambda^2)$; d is the dimensionality, and is equal to 3 for the studied point defects; $E_m$ is the migration barrier presented in Table I; $k_B$ is the Boltzmann's constant; T is the temperature. $D_0$ is the pre-exponential constant of diffusivity, which is also shown in Table I.

*2. Transformation of Si vacancies*

It has been found that the Si vacancy can transform into a $V_C$-$C_{Si}$ complex,[15,16] with a barrier of 2.5eV~2.7eV for n-type cubic SiC. Using density functional theory (DFT) with the GW approximation, Bruneval et al.[47] has revealed that the $V_{Si}$ appears to be metastable with respect to the $V_C$-$C_{Si}$ complex for any Fermi level in the band gap for 3C-SiC, and found that a transformation occurs with a lower barrier of 2.32eV from $V_{Si}^-$ to $(V_C\text{-}C_{Si})^-$ mediated by a charge change from $V_{Si}^-$ to $V_{Si}^+$, while a direct transformation from $V_{Si}^-$ to $(V_C\text{-}C_{Si})^-$ requires a higher barrier of 2.75eV. As a first attempt, a barrier of 2.7eV for the $V_{Si}$ transformation, and a value of 2.4eV for the reverse transformation from $V_C\text{-}C_{Si}$ to $V_{Si}$, used in the published ab-initio based rate theory calculation,[29] were employed in this study. The transformation rate is calculated as:

$$v_t = v_t^0 \exp(-\frac{E_t}{k_B T}) \quad (2)$$

where $v_t^0$ is as assumed to be the same as that of $V_{Si}$ migration, on the order of $10^{11}$Hz. A variation from $10^{11}$ Hz to $10^{13}$ Hz does not have significant effects on the results.

*3. Reactions between point defects*

Reactions considered in this study include recombination reactions and kick-out reactions. Recombination between interstitials and vacancies can be either homogeneous or heterogeneous, with the former resulting into complete recovery and the latter producing an antisite defect. The kick-out reactions can partially heal the damage and produce a mobile interstitial for further reactions. Interstitials can also recombine with $V_C\text{-}C_{Si}$ complexes, producing an antisite or an antisite pair.[52]

Depending on the migration barriers of point defects and the reaction barriers of their reactions, point defect reactions can have different energy landscapes (ELs). The complexities of ELs in SiC have been indicated in Ref. 17 and have been found to be important for the long term defect evolution.[29] Based on the energies of point defects in Table I and the parameters in Table II, point defect reactions are treated by the methods described in Ref. 32.

For the reactions to occur, the reactants have to approach each other within a reaction distance (r) to form a defect complex. If the reaction barrier of the complex ($E_{Re}$) is smaller than the migration barrier of the fastest reactants ($E_m$), we assume a diffusion limited reaction (R1, R2, R4, R6), and the reaction occurs immediately. When $E_{Re}$ is higher than $E_m$, the reaction can be of either recombination EL or trapping EL.

For reactions of recombination EL, including R3, R7 and R8 in Table II, the defect complex can dissociate with a rate $v_d$, depending on the binding energy ($E_B$) of the complex and the migration barrier of the fastest reactant ($E_m$). In addition, the complex can also overcome the reaction barrier ($E_{Re}$) and react with a rate $v_R$:

$$v_d = v_d^0 \exp(-\frac{E_B + E_m}{k_B T}) \quad (3)$$

$$v_R = v_R^0 \exp(-\frac{E_{Re}}{k_B T}) \quad (4)$$

Where $v_d^0$ and $v_r^0$ are the attempt frequencies of dissociation and recombination, and are approximated as that of migration ($v_m^0$) for the fastest reactant. The energy landscape of the kick-out reaction R5 has been found to be a trapping profile.[17] The defect complex can recombine with a rate $v_R$ calculated by Equation (4), while its dissociation rate is calculated as:

$$v_d = v_d^0 \exp(-\frac{E_{Tr}}{k_B T}) \quad (5)$$

Here $E_{Tr}$ is the trapping barrier for the defect complex.

*4. Recombination of antisite pair defects*

The nearest-neighbor antisite pair defects, denoted as $Si_C\text{-}C_{Si}$, can be detected in 3C-SiC after electron and proton irradiation.[53] MD simulations have revealed that antisite pairs can be introduced by displacement cascades.[54] As mentioned above, recombination between $Si_I$ and $V_C\text{-}Si_C$ complexes can also result in antisite pairs.[52] Besides these two origins, some mechanisms for the defect formation have been proposed, including a kick-out mechanism by carbon and silicon interstitials;[55] a mechanism involving vacancy migration and vacancy-assisted motion of antisites;[51] and a low energy mechanism by Si atoms not completely removed from its lattice site.[53] However, for simplicity, only antisite pair defects introduced directly by MD displacement cascades and by recombination of $Si_I$ and $V_C\text{-}Si_C$ are considered in this study. The antisite pair defects have been found thermally unstable and can recombine with a barrier of 0.65~3.2eV.[56-58] In this study, the rate for the recovery of antisite pair defects is estimated by $v_0\exp(-3.2eV/k_B T)$,[58] with $v_0$ a value of

**Table II** Binding energies ($E_B$)/trapping barriers ($E_{Tr}$), reaction barrier ($E_{Re}$) and reaction distances (r) for point defect reactions

| Reaction | $E_B$ or $E_{Tr}$ (eV) | $E_{Re}$(eV) | r (Å) |
|---|---|---|---|
| R1: $C_I+V_C \leftrightarrow C_I\text{-}V_C \rightarrow C_C$ | 1.69[a] | 0.43[a] | 3.08[a] |
| R2: $Si_I+V_{Si} \leftrightarrow Si_I\text{-}V_{Si} \rightarrow Si_{Si}$ | 0.13[a] | 0.17[a] | 5.34[a] |
| R3: $C_I+V_{Si} \leftrightarrow C_I\text{-}V_{Si} \rightarrow C_{Si}$ | 2.96[b] | 1.25[b] | 3.30[b] |
| R4: $Si_I+V_C \leftrightarrow Si_I\text{-}V_C \rightarrow Si_C$ | 0.50[b] | 1.11[b] | 3.70[b] |
| R5: $C_I+Si_C \leftrightarrow C_I\text{-}Si_C \rightarrow C_C+Si_I$ | 1.67[a] | 1.34[a] | 4.36[a] |
| R6: $Si_I+C_{Si} \leftrightarrow Si_I\text{-}C_{Si} \rightarrow Si_{Si}+C_I$ | -0.33[a] | 0.64[a] | 4.36[a] |
| R7: $C_I+V_C\text{-}C_{Si} \leftrightarrow C_I\text{-}V_C\text{-}C_{Si} \rightarrow C_C+C_{Si}$ | 1.39[b] | 0.8[b] | 4.70[b] |
| R8: $Si_I+V_C\text{-}C_{Si} \leftrightarrow Si_I\text{-}V_C\text{-}C_{Si} \rightarrow Si_C\text{-}C_{Si}$ | 0.73[b] | 3.12[b] | 4.40[b] |

[a]Ref. 17; [b]Ref. 52;

$10^{13}$Hz.

*5. Clustering between point defects and the growth and the shrinkage of defect clusters*

In this study, a defect cluster is defined as a group of vacancies (interstitials) situated within the 2NN distance.[27] According to previous DFT studies[17,43] and a recent DFT calculation with a hybrid functional,[59] the stable charge states for point defects are neutral except for the negatively charged $V_{Si}$, for n-type doping 3C-SiC. To account for $V_{Si}$, in our model, we assume that $V_{Si}$ will not cluster with each other during OKMC simulation due to the short-range electrostatic repulsion. This assumption should be reasonable because the binding energy for di-$V_{Si}$ cluster is relatively low (0.1~0.21eV).[60]

Previous ab-initio studies on aggregation of $C_I$ have demonstrated that $C_I$ can be trapped by $C_{Si}$ and form stable carbon clusters with the configuration of di-carbon antisite.[61-63] Though there has been no evidence showing that $Si_I$ can be captured by antisites or $C_I$ can aggregate at $Si_C$, we assume that interstitials can form defect clusters with antisite defects with a capture distance of 2NN. The resulting clusters are also referred as interstitial clusters in this study. The shape of vacancy clusters is considered to be irregular, depending on the positions of their constituent vacancies.[32] Considering that the ground state (GS) configurations of clusters with up to 6 carbon atoms are mostly parallel to the {111} plane,[63] we model them as disc-shaped planar defects parallel to the plane.

Since antisite defects are almost immobile below 1200K, with migration barriers above 10eV, defect clusters are limited to react with interstitials and vacancies only. The interstitial (vacancy) clusters will grow by absorbing new coming interstitials (vacancies), and shrink by vacancies (interstitials) recombination. Such growth and shrinkage are assumed to occur instantaneously within a capture radius of 2NN.

Regarding the dissociation of interstitial clusters, we assume that the interstitial clusters are thermally stable and will not emit point defects below 1200K. This assumption could be safe for carbon clusters with size smaller than 4,[62-64] and Si-C clusters[65] due to the reported high dissociation barriers (higher than 4.0eV). For Si clusters, Hornos et al.[65] have reported a binding energy of 3.16eV (with a dissociation barrier higher than 4.5eV) for di-Si interstitial clusters, which seems stable for this study. However, Liao et al.[66] have recently reported a lower binding energy of 1.98eV for the di-Si clusters. In addition, a recent ab-inito study by H. Jiang et al.[64] has shown that the dissociation barrier for carbon clusters exhibits a tendency to decrease with the increase of cluster size. Their results could imply that interstitial clusters can become unstable and emit interstitials as its size gets large enough. The impact of this assumption on defect annealing will be discussed in Section III(A).

Although we have assumed that $V_{Si}$ cannot aggregate into $V_{Si}$ clusters during OKMC simulations, the $(V_{Si})_n$ clusters can be pre-implanted directly by MD cascades. Indicated by the low binding energy of di-$V_{Si}$ clusters and the negative charge state of $V_{Si}$, the binding between $V_{Si}$ in the clusters should be weak. As a result, the $(V_{si})_n$ clusters are assumed to be unstable and can dissociate with a binding energy of 0eV. The binding energies of di-$V_C$ clusters,[16,60] and clusters formed by carbon and silicon vacancies ($nV_C$-$V_{Si}$ and $V_C$-$nV_{Si}$)[60,67] have been calculated to be about 1eV, and higher than 2eV, respectively. Adding the migration barrier of carbon or silicon vacancies leads to high dissociation barriers (>4.5eV). Though it has been reported that the binding energies of vacancy clusters decrease as the cluster size increases,[60] the $(V_C)_n$ and the $nV_C$-$mV_{Si}$ clusters are assumed to be stable and not to emit vacancies as a first attempt.

*6. Setup for OKMC simulations*

In the present OKMC calculations, a simulation volume of 143.5nm × 143.5nm × 143.5 nm ($329a_0 \times 329a_0 \times 329a_0$) has been used, with periodic boundaries applied in all directions. Cascades debris from MD simulation and uncorrelated isolated point defects were introduced randomly in the simulation box at 160K, where the dynamical annealing is insignificant and the dose-rate effects can be ignored.[68] For the case of cascade damage annealing, the total damage is considered linearly decomposed into a number of 10keV MD cascades, with the number of 10keV cascade determined by $N_{10keV} = \frac{E_{dam}}{10keV}$. Here, $E_{dam}$ is the damage energy corresponding to the dose, calculated by SRIM with the procedures depicted in Section II(A). Regarding the case of annealing the randomly distributed point defects, N.Swaminathan et al.[30] have calculated the number of point defect by electron irradiation using $N_i = \Gamma \eta \alpha_n$ (per atom), where $\Gamma$ is the dose (dpa); $\eta$ is the cascade efficiency and is taken as 0.8;[30] $\alpha_n$ are the fractions of defect of type i, determined in MD simulation of low energy cascades[69] as: $\alpha_{Vc}$ =0.330; $\alpha_{Ic}$ =0.380; $\alpha_{Vsi}$ =0.084; $\alpha_{SiI}$ =0.035; $\alpha_{Csi}$ =0.063; $\alpha_{Sic}$ =0.110. In their definition, the "displacement" of "dpa" (displacement per atom) was referred as all the defects including interstitials, vacancies and antisites. However, for dpas calculated by the NRT approach, the "displacement" should represent the number of events where an atom is displaced from its lattice site and an interstitial-vacancy pair (Frenkel pair) is created. For better consistency with the dose of ion irradiations estimated by NRT calculations, we estimated the number of point defects by $N_i = \Gamma \eta \frac{\alpha_i}{\alpha_{Si_I} + \alpha_{C_I}}$, with $\alpha_i$, $\Gamma$ and $\eta$ the same as those of N.Swaminathan et al.[30]

## II. Calculation approach
### A. Comparison between simulations and experiments

Based on the model and the parameters shown in Section II, the experiment of 1nm$^{-2}$ 550keV Si$^+$ implantation was reproduced by OKMC simulations. MD cascades of 10keV Si PKAs were introduced randomly into the volume with a dose rate of 2.3×10$^{-4}$dpa/s up to a dose of 0.023dpa at 160K. Subsequently, the whole system was annealed isochronally from 160K to 1170K, with each annealing step lasting 1200 s.

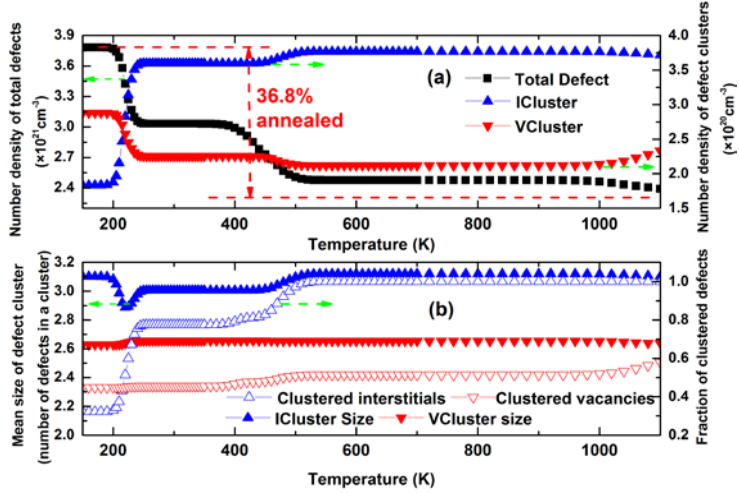

FIG. 3. (a) Number density of total defects and defect clusters; (b) fraction of clustered point defects and the mean size of defect clusters as a function of temperature during the isochronal annealing.

RBS/C is a useful technique to determine the lattice disorder on the Si sub-lattice in SiC.[10,34] A previous study on disorder accumulation and recovery in 3C-SiC[34] has revealed that disorder on Si and C sub-lattices have similar recovery behavior along multiple directions, which indicates that the total disorder on both sub-lattices should follow a similar trend as that on the Si sub-lattice. The disorder measured by the ion channeling techniques consists of contributions from amorphous materials, interstitials and small interstitial clusters.[70] Therefore, the annealing behavior of total interstitials, including clustered or isolated ones, can be compared to the disorder on the Si sub-lattice observed by RBS/C. To compare the annealing behavior, the relative fraction of annealing was introduced as:

$$R(T) = \frac{(N(T_{min}) - N(T))}{(N(T_{min}) - N(T_{max}))} \quad (6)$$

Where $N(T_{min})$ and $N(T_{max})$ are the number of defects at the minimum and maximum studied temperature and $N(T)$ is the value at a specific temperature $T$. The relative fraction of annealing defines the accumulative fraction of annealing from $T_{min}$ to $T$ relative to the total annealing from $T_{min}$ to $T_{max}$, and reflects the annealing behavior within the studied temperature range.

The relative fraction of annealing for the RBS disorder, derived from the literature[10] and that of total interstitials from OKMC simulation were calculated with Equation (6). The results are shown in Fig. 2.

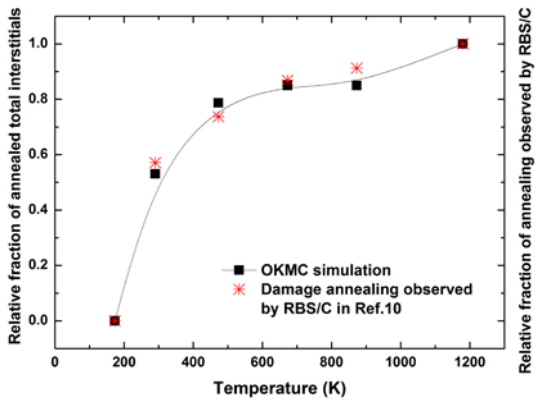

FIG.2. Relative fraction of annealing for disorder observed by RBS/C from Ref.10 vs total interstitial in this study

As shown in the Fig. 2, the annealing behavior of total interstitials in our simulation is in agreement with that observed by RBS/C, which validates the current parameterization. Some small discrepancies can be observed at 670K~870K. Within the temperature regime, almost no annealing occurs in the simulation, while 5% of the total annealing is observed by RBS/C experimentally. In the previous experiments on isochronal annealing of Al-implanted SiC, recovery between 600K~800K has been observed.[12] Such annealing stage is absent at low ion fluencies, which are close to the fluence in this study, and only becomes considerable at intermediate and high fluencies. As indicated in this study, the mobility of vacancies is limited and all the interstitials are contained in the interstitial clusters (in Section III(B)) within the temperature range. Such annealing stage is most probably attributed to the dissociation of interstitial clusters. When the ion fluence is high enough, large interstitial clusters can be produced. As the dissociation barriers of interstitial clusters decrease with the increase of cluster sizes, the large clusters can dissociate and emit interstitials at high temperature, enhancing the total interstitials. Therefore, the discrepancies in the annealing of RBS disorder and total interstitials originate from the assumption that the dissociation of interstitial clusters is ignored in this study. It can be concluded that our current model is suitable for simulating recovery of damage by low-fluence ion irradiation. However, for high doses at high temperatures, dissociation of large interstitial clusters should be considered in future studies.

**B. Recovery during isochronal annealing**

To further investigate the mechanism of damage recovery, using the same implantation setup as in Section III(A), a continuous isochronal annealing was carried out from 160K to 1100K with temperature intervals of 5 to 20K and isochronal steps of $\Delta t=300s$. The evolution of the number density of total defects as well as the clustering as a function of temperature is shown in Fig. 3.

As presented in Fig. 3(a), it can be found that only 36.8% of the total defects are annealed at 1100K. The result is similar to the incomplete annealing of cascade damage in 3C-SiC observed in the previous kinetic lattice Monte Carlo (KLMC) modeling,[27] where more than half of the original defects still remain after annealing at 1100K. Resistivity measurements of neutron-irradiated 3C-SiC have shown that 90% of irradiation defects can be removed by annealing at 350°C for 5 min.[71] However, it is not straightforward to quantify the annealing of total defects by the recovery of resistivity, because the contributions by different defects to the free-carrier removal are unclear. The incomplete annealing has also been demonstrated by experiments showing that thermally stable defects, like $D_I$ and $D_{II}$ PL centers, can be formed during the irradiation of energetic particles and can persist up to 2000K.[72,73] The inefficient annealing observed in this study is due to the significant clustering of interstitials as indicated in Fig. 3(b). The clustering fraction of interstitial reaches 77% at 260K, and above 540K all remaining interstitials are confined in interstitial clusters. The interstitials clustered into immobile defects cannot be involved in recombination unless the vacancies can migrate, which requires high temperature. The number density of interstitials as well as the fraction of clustered interstitials increases at 200K~245K and 450K~510K, which corresponds to the migration of $C_I$ and $Si_I$. In addition, small decreases in the number density and the size of interstitial clusters occur due to recombination between migrating vacancies and the clusters at high temperature. Fig.3(b) shows an average size of n=3.1 for interstitial clusters at the beginning of annealing. The subsequent decrease of cluster sizes at low temperature is attributed to the fast production of di-interstitial clusters at this temperature regime. The variation in the fraction of clustered vacancies is less than that of interstitials, owing to a much lower mobility of vacancies. The increase in the clustering fraction of vacancies at about 400K arises from the recombination between un-clustered vacancies and other point defects, while the increase above 900K is caused by the migration of vacancies. The mean size of vacancy clusters remains almost unchanged during the annealing. Regarding the number density of vacancy clusters, the decrease at low temperature results from the recombination between migrating interstitials and the vacancy clusters. The number density re-grows as vacancies begin to migrate at high temperature.

From Fig. 3(a), it can be found that there are several recovery stages during isochronal annealing. The differential isochronal recovery spectrum derived from Fig. 3(a) and the evolutions of different defects are combined in Fig. 4 to interpret the mechanism of those annealing stages.

Three main annealing stages can be identified in Fig. 4(a): (I) 200~250K; (II) 400~520K; (III) >900K. Stage I and Stage II in our model agree with the two annealing stages below 600K identified in the experiment:[11] (I) 150~300K; (II) 450~550K. The stage observed between 570K and 720K in Ref. 11, which exists under high fluence irradiation, is absent in the study. As indicated in the previous Sec. III A, annealing between 600~800K is attributed to the dissociation of large interstitial cluster. However, such process is insignificant under the studied damage dose, as is shown in Fig. 2 that annealing in this stage accounts for only 5% of the total recovery, and was absent in this current study. It is expected that a less prominent recovery peak could appear at 600~800K if the dissociation of interstitial cluster was considered.

Fig. 4(b) shows the concentration of different defects as a function of annealing temperature. Combined with Fig. 4(a) and (b), mechanisms responsible for each annealing stage are presented as follow:

1. Stage I corresponds to the recombination of $C_I$ with $V_C$ and vacancy clusters (VClusters) due to the migration of $C_I$. The increase of $V_{Si}$ in this stage is caused by the $V_{Si}$ remaining in the recombination of $C_I$ with $V_{Si}$-$nV_C$ clusters. As opposed to the annealing of iron, where the first recovery stage can be divided into two sub-stages (a more prominent $I_D$ for close-pair and correlated recombination and a less prominent $I_E$ for uncorrelated recombination),[22,24] there is only one peak for Stage I of SiC. As the recombination barrier of $C_I$ and $V_C$ is lower than the migration barrier of $C_I$, close-pair $C_I$-$V_C$ can recombine before the migration of $C_I$. Though not shown here, the close-pair recombination does exist, yet such recombination is negligible. The difference between the damage recovery in metals and SiC can be due to the more dispersed structure of SiC cascades.[18,19]

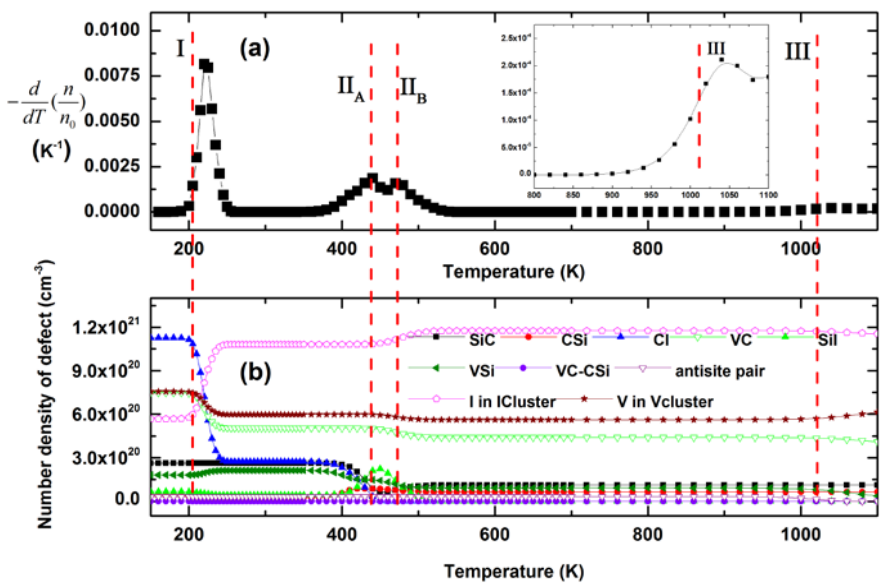

FIG. 4. (a) Differential isochronal recovery spectrum, where n denotes the total defect at temperature T, and $n_0$ represents the total defect at the beginning of annealing. (b) Evolution of different types of defects during isochronal annealing. The dash lines indicate the position of the recovery stages.

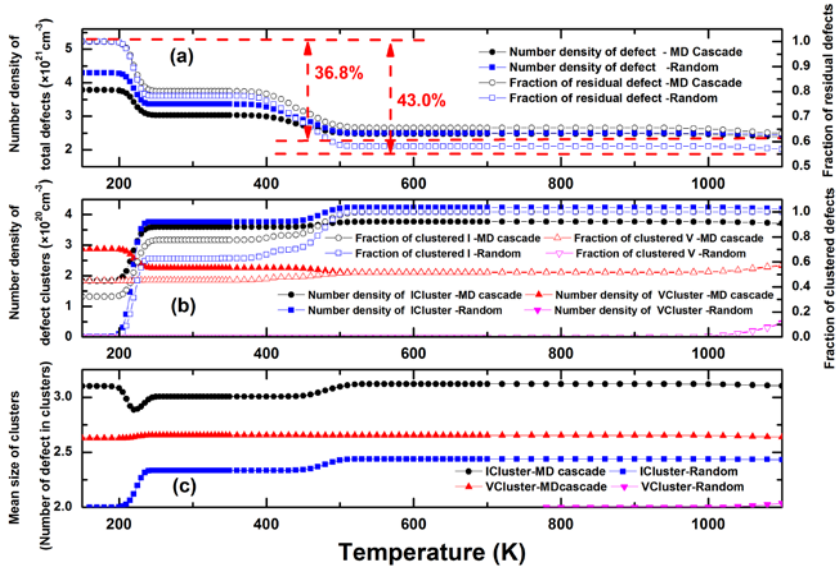

FIG. 5.(a) Number density of total defect and fraction of residual defects; (b) Number density of interstitial clusters (ICluster) and vacancy clusters (VCluster) and fraction of clustered point defects; (c) Mean size of interstitial clusters and vacancy clusters, during the isochronal annealing of cascade damage and uniformly distributed isolated point defects.

2. Stage II includes two recovery peaks. Stage $II_A$ starts at 400K, and is caused by the heterogeneous recombination between $C_I$ and $V_{Si}$, and the kick-out reaction between $C_I$ and $Si_C$ leading to the increase of $C_{Si}$ and $Si_I$ below 450K. Stage $II_B$ dominates above 460K, when the migration of $Si_I$ becomes significant. At stage $II_B$, the annealing is attributed to the recombination of $V_{Si}$ and VClusters with $Si_I$, the heterogeneous recombination between $Si_I$ and $V_C$ which leads to a slight re-growth of $Si_C$, and the kick-out reaction between $Si_I$ and $C_{Si}$.

3. Stage III is much less prominent compared to the previous two stages. Fig. 4(b) shows that this recovery stage corresponds to the annihilation of $V_{Si}$ at interstitial clusters (IClusters), due to the migration of $V_{Si}$, and the recovery of antisite pair defects. The decrease at higher temperature seen in the inset of Fig. 4(a) arises from the annealing-out of the antisite pairs.

## C. Influence of spatial correlation in primary damage

To study the influence of spatial correlation of primary damage structure, randomly distributed isolated point defects and MD cascades were implanted with the same conditions (dose, dose rate, temperature). The cascade damage has the typical morphology of damage by ion implantation and neutron irradiation, where defect clusters are directly introduced into the cascades, and some of the defects are spatially correlated. In the case of randomly distributed isolated point defects, the spatial correlation between defects is removed, which mimics the case of mean field rate theory (MFRT) calculations. In fact, it has been shown that when defects are uniformly distributed, KMC and MFRT models are in reasonable agreement.[74] In this section, defects in the form of cascades and isolated defects were annealed isochronally with the temperature and time steps adopted in Section III(B).

Fig. 5 gives the evolution of number density of total defects and defect clusters, as well as the defect cluster size, as a function of annealing temperature, both with a dose of 0.023dpa.

As shown in Fig. 5(a), the initial concentration of total defects of the displacement cascades is lower by 13.4% than that of the isolated point defects, probably due to higher intra-cascade recombination of cascade damage during implantation. Compared to the case of cascade damage, the fraction of annealing is about 7% higher in the case of uniform isolated defects, because the fraction of clustered interstitial is lower at low temperature shown in Fig. 5(b), leaving more mobile interstitials for recombination. The difference in the annealing behaviors of cascade damage and isolated point defects of SiC is much less significant than that of Fe,[23] where most of the defects have been annealed by 500K for isolated defects, but the annealing of cascade damage was insignificant up to that temperature. The less significant difference in SiC compared to iron can be attributed to the fact that SiC cascades possess dispersed structures with low density multiple branches,[18,19] while the Fe cascades have compact spherical structures with vacancy-rich cores surrounded by interstitial-rich regions,[18,75] much more different from uniformly distributed point defects. The number density of defect clusters, as well as the fraction of clustered defects for the cases of cascade and random damage is compared in Fig.5(b). The number density of interstitial clusters of random damage grows from zero, and surpasses that of cascade damage above 250K. However, revealed by Fig. 5(c), the average size of interstitial clusters of random damage remains smaller than that of cascade damage during the annealing. The clustering of vacancies is much less significant in the case of random damage, where vacancy clusters can only be detected at high temperature and the average size of vacancy clusters is smaller, compared to the case of cascade damage.

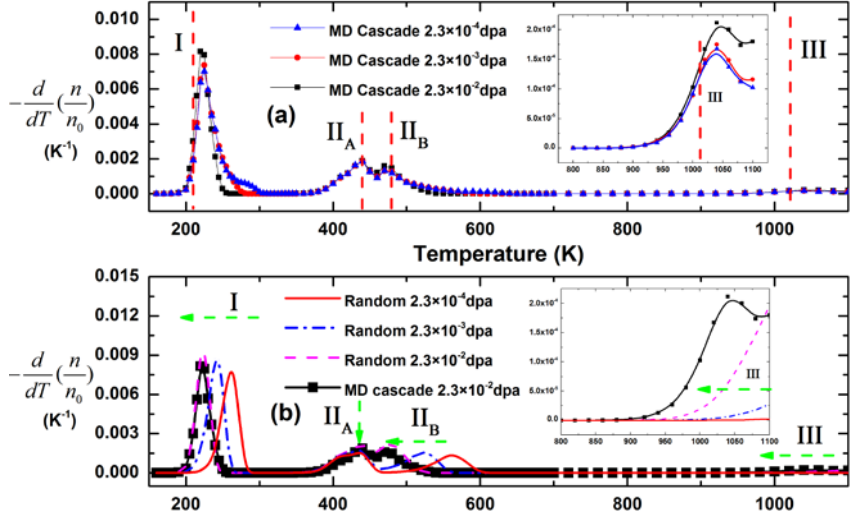

FIG. 6. Differential isochronal recovery spectra of (a) cascade damage; (b) isolated point defects, where the dashed lines denote the position of the recovery stages and the dashed arrows indicates the shifting of the stages.

As mentioned in the previous Sec. III B, three recovery stages are identified within the studied temperature regime. To study the effects of spatial correlation in primary damage on each recovery stage, differential isochronal recovery spectra of cascade damage and isolated point defects are compared. In addition, since the damage dose can influence the mean I-V distance and the distance between displacement cascades, the effects of dose should be considered when studying the effects of spatial correlation. Therefore, the results of cascade damage and isolated point defects with doses from $2.3 \times 10^{-4}$ dpa to $2.3 \times 10^{-2}$ dpa are presented in Fig. 6.

For the annealing of cascade damage, the increase in dose only leads to variations in the width and height of recovery peaks, without shifting the position of peaks with temperature, which is similar to Fe cascades.[23] The lower end of temperature at Stage I corresponds to the onset of $C_I$ migration, which only depends on the migration barriers and does not change with dose. However, the width of the recovery peak shrinks with increase in dose. At the lowest dose, when cascades are well separated, inter-cascade recombination can occur when interstitials migrate and recombine with vacancies in other cascades, leading to a shoulder at about 300K. As the dose increases, the separation between cascades becomes smaller, reducing the migration distance for the inter-cascade recombination and making the shoulder smaller. However, the in-cascade recombination is unaffected and the width of the main peak remains unchanged. At the highest dose, cascade overlap begins and the in-cascade recombination occurs with shorter mean I-V distances, leading to the narrowing of the recovery peak. Similar mechanisms for peak narrowing can be applied to Stage $II_B$ and Stage III which require migration of point defects. However, the peak narrowing is to some extent insignificant for Stage III, as reactions of antisite pair defects, whose occurrence does not depend on defect migration, contribute to defect recovery in this stage. Stage $II_A$ is not influenced by the dose, as it is caused by the reaction of defect complexes formed at Stage I or from the initial MD cascades.

As shown in Fig. 6(b), the annealing of isolated point defects is much affected by the dose, which is similar to the modeling of electron irradiation in Fe where only isolated point defects are introduced.[22] As the dose increases, recovery peaks except that of Stage $II_A$ shifts to lower temperatures, due to the shorter mean migration distance needed by point defects to recombine, while the position of the recovery peak $II_A$ does not change as in the annealing of cascade damage.

To compare the recovery stages of cascade damage and isolated point defects, the annealing of 0.023dpa cascade damage is also reproduced in the Fig. 6(b). The peak corresponding to the recovery of antisite pairs does not exist in Stage III due to the absence of the defects. However, it is interesting to find that at the highest dose, the recovery stages of cascade damage resemble those of isolated point defects, possibly because the mean I-V distance in the case of isolated defect with the highest dose is close to that within displacement cascades. The results indicate that the spatial correlation of primary damage structure can be unimportant for the annealing behavior of radiation damage in SiC, when the dose is high.

However, in this study, the dissociation of interstitial clusters was ignored, which is expected to play a role for high dose damage at high temperature. As shown in Fig.5(c), the mean size of defect clusters in the case of cascade damage is larger than that of random isolated point defects. Considering that the decreasing tendency of dissociation barriers with the increase of cluster sizes indicated by H. Jiang et al.,[64] the larger clusters in cascade damage are more inclined to dissociate, as it is the case of Fe where additional recovery peaks related to the dissociation of large clusters can be found in the annealing of cascade damage.[23] In addition, the number density of vacancy cluster is much lower, and the annealing of antisite pair defects is absent in the case of isolated point defects. Such effects can be mitigated by introducing defect clusters and antisite pairs randomly in the case of isolated defects.

## IV. Conclusions

Based on a framework of multi-scale modeling, we have studied the long-term evolution of displacement damage induced by heavy ion irradiation in SiC. The experiment on isochronal annealing of heavy ion irradiated SiC has been simulated and the results on the annealing behaviors of total interstitials show good agreement with previous experimental results. The clustering of interstitials is significant, which hinders the complete recovery of defects. Studies on the mechanisms of defect recovery reveal two annealing stages below 600K, in agreement with reported experiments. The first stage occurs near the room temperature and originates from the recombination of $C_I$ with $V_C$ and vacancy clusters, due to the migration of $C_I$. Opposed to the case of iron, close-pair recombination is found to be negligible in this stage, possibly due to the dispersed nature of SiC cascades. The second stage is related to a number of processes, and can be divided into two sub-stages. The first sub-stage results from the heterogeneous recombination between $C_I$ and $V_{Si}$, and the kick-out reaction between $C_I$ and $Si_C$. In the second sub-stage, the $Si_I$ is mobile, which leads to the homogeneous and heterogeneous recombination of $Si_I$ with $V_{Si}$ and $V_C$, and the kick-out reaction between $Si_I$ and $C_{Si}$. An extra annealing stage is identified at temperature higher than 900K, which results from the annihilation of $V_{Si}$ at interstitial cluster, due to the migration of $V_{Si}$, and the recovery of antisite pair defects. The fraction of annealing is about 7% higher when implanted defects are uniformly isolated, compared to the case of cascade damage. At the highest dose of this study, the annealing behavior of isolated point defects and cascade damage is found to be similar, which indicates that the influence of spatial correlation in primary damage is insignificant for SiC, when the damage dose is high enough. Combining such weak influence of spatial correlation and the insignificant close-pair recombination in the first recovery stage, and under the condition that the in-cascade clustering is considered, it is suggested that approaches like MFRT with mean-field approximation can simulate the long-term evolution of high dose cascade damage in SiC.

## Acknowledgments

This work was supported by the National Natural Science Foundation of China (Grant No. 11175138), the National Science Foundation for Post-doctoral Scientist of China (Grant No. 2013M540757), the Key Program of the National Natural Science Foundation of China (Grant No. 11235008), and the State Key Laboratory of Intense Pulsed Radiation Simulation and Effect (Grant No. 20140134).

I. Martin-Bragado wants to acknowledge partial funding from the project MASTIC (PCIG09-GA-2011-293783) by the Marie Curie actions grant FP7-PEOPLE-2011-CIG program and also from the Spanish Ministry of Economy and Competitiveness under Ramon y Cajal fellowship RYC-2012-10639. The work contributes to the Joint Programme of Nuclear Materials of the European Union Energy Research Alliance.

D. Guo appreciates the support by the National Science Foundation of China (No. 11305122), China Postdoctoral Science Foundation (2013M532052), and the Fundamental Research Funds for the Central University (xkjc2014009). Molecular dynamics simulations were performed on the cluster Hua-I in Xi'an Jiaotong University.

We would like to thank Dr. G. Roma from CEA for useful discussions.

## References

1) W. Wondrak, R. Held, E. Niemann, and U. Schmid, Ieee Transactions on Industrial Electronics **48,** 307 (2001).
2) P. Godignon, X. Jorda, M. Vellvehi, X. Perpina, V. Banu, D. Lopez, J. Barbero, P. Brosselard, and S. Massetti, Ieee Transactions on Industrial Electronics **58,** 2582 (2011)
3) G. Izzo, G. Litrico, L. Calcagno, G. Foti, and F. La Via, Journal of Applied Physics **104,** 093711 (2008)
4) A. Hallen, M. Nawaz, C. Zaring, M. Usman, M. Domeij, and M. Ostling, Ieee Electron Device Letters **31,** 707 (2010)
5) C. Brisset, O. Noblanc, C. Picard, F. Joffre, and C. Brylinski, Ieee Transactions on Nuclear Science **47,** 598 (2000)
6) J. P. Crocombette, G. Dumazer, N. Q. Hoang, F. Gao, and W. J. Weber, Journal of Applied Physics **101,** 023527 (2007).
7) J. Cabrero, F. Audubert, R. Pailler, A. Kusiak, J. L. Battaglia, and P. Weisbecker, Journal of Nuclear Materials **396,** 202 (2010).
8) W. Jiang, C. M. Wang, W. J. Weber, M. H. Engelhard, and L. V. Saraf, Journal of Applied Physics **95,** 4687 (2004).
9) H. Z. Xue, Y. Zhang, Z. Zhu, W. M. Zhang, I. T. Bae, and W. J. Weber, Nuclear Instruments & Methods in Physics Research Section B-Beam Interactions with Materials and Atoms **286,** 114 (2012).
10) W. J. Weber, W. Jiang, and S. Thevuthasan, Nuclear Instruments & Methods in Physics Research Section B-Beam Interactions with Materials and Atoms **166,** 410 (2000).
11) W. J. Weber, W. Jiang, and S. Thevuthasan, Nuclear Instruments & Methods in Physics Research Section B-Beam Interactions with Materials and Atoms **175,** 26 (2001).
12) Y. Zhang, W. J. Weber, W. Jiang, A. Hallen, and G. Possnert, Journal of Applied Physics **91,** 6388 (2002).
13) F. Gao and W. J. Weber, Journal of Applied Physics **94,** 4348 (2003).
14) F. Gao, W. J. Weber, M. Posselt, and V. Belko, Physical Review B **69,** 245205 (2004).
15) M. Bockstedte, A. Mattausch, and O. Pankratov, Physical Review B **68,** 205201 (2003).
16) M. Bockstedte, A. Mattausch, and O. Pankratov, Physical Review B **69,** 235202 (2004).
17) M. J. Zheng, N. Swaminathan, D. Morgan, and I. Szlufarska, Physical Review B **88,** 054105 (2013).
18) R. Devanathan, W. J. Weber, and F. Gao, Journal of Applied Physics **90,** 2303 (2001).
19) F. Gao and W. J. Weber, Physical Review B **63,** 054101 (2000).
20) D. E. Farrell, N. Bernstein, and W. K. Liu, Journal of Nuclear Materials **385,** 572 (2009).
21) H. X. Xu, Y. N. Osetsky, and R. E. Stoller, Journal of Nuclear Materials **423,** 102 (2012).
22) C. C. Fu, J. Dalla Torre, F. Willaime, J. L. Bocquet, and A. Barbu, Nature Materials **4,** 68 (2005).
23) C. J. Ortiz and M. J. Caturla, Physical Review B **75,** 184101 (2007).
24) R. Ngayam-Happy, P. Olsson, C. S. Becquart, and C. Domain, Journal of Nuclear Materials **407,** 16 (2010).
25) M. Hou, A. Souidi, C. S. Becquart, C. Domain, and L. Malerba,

25) Journal of Nuclear Materials **382,** 103 (2008).
26) A. Souidi, M. Hou, C. S. Becquart, L. Malerba, C. Domain, and R. E. Stoller, Journal of Nuclear Materials **419,** 122 (2011).
27) Z. Rong, F. Gao, W. J. Weber, and G. Hobler, Journal of Applied Physics **102,** 103508 (2007).
28) R. E. Stoller, S. I. Golubov, C. Domain, and C. S. Becquart, Journal of Nuclear Materials **382,** 77 (2008).
29) N. Swaminathan, D. Morgan, and I. Szlufarska, Journal of Nuclear Materials **414,** 431 (2011).
30) N. Swaminathan, D. Morgan, and I. Szlufarska, Physical Review B **86,** 214110 (2012).
31) J. Dalla Torre, C. C. Fu, F. Willaime, A. Barbu, and J. L. Bocquet, Journal of Nuclear Materials **352,** 42 (2006).
32) I. Martin-Bragado, A. Rivera, G. Valles, J. L. Gomez-Selles, and M. J. Caturla, Computer Physics Communications **184,** 2703 (2013).
33) Modular Monte Carlo (MMonCa) see: http://www.materials.imdea.org/MMonCa/
34) W. Jiang, W. J. Weber, J. Lian, and N. M. Kalkhoran, Journal of Applied Physics **105,** 013529 (2009).
35) J. F. Ziegler, Nuclear Instruments & Methods in Physics Research Section B-Beam Interactions with Materials and Atoms **219,** 1027 (2004).
36) F. Gao, W. J. Weber, and R. Devanathan, Nuclear Instruments & Methods in Physics Research Section B-Beam Interactions with Materials and Atoms **191,** 487 (2002).
37) R. E. Stoller, M. B. Toloczko, G. S. Was, A. G. Certain, S. Dwaraknath, and F. A. Garner, Nuclear Instruments & Methods in Physics Research Section B-Beam Interactions with Materials and Atoms **310,** 75 (2013).
38) Large-scale Atomic/Molecular Massively Parrlel Simulator (LAMMPS) see: http://lammps.sandia/gov/
39) J. Tersoff, Physical Review Letters **61,** 2879 (1988).
40) R. Devanathan, W. J. Weber, and T. D. de la Rubia, Nuclear Instruments & Methods in Physics Research Section B-Beam Interactions with Materials and Atoms **141,** 118 (1998).
41) A. F. Voter, in *Radiation Effects in Solids*, edited by K. E. Sickafus. and E. A. Kotomin (Springer, NATO Publishing Unit, Dordrecht, The Netherlands, 2005).
42) F. Bernardini, A. Mattoni, and L. Colombo, European Physical Journal B **38,** 437 (2004).
43) D. Shrader, S. M. Khalil, T. Gerczak, T. R. Allen, A. J. Heim, I. Szlufarska, and D. Morgan, Journal of Nuclear Materials **408,** 257 (2011).
44) E. Rauls, Z. Hajnal, A. Gali, P. Deak, and T. Frauenheim, Mater. Sci. Forum **353-356,** 435 (2001).
45) E. Rauls, "Annealing mechanisms of point defects in silicon carbide", Ph. D. thesis (University of Paderbon, 2003).
46) S. Kondo, Y. Katoh, and L. L. Snead, Physical Review B **83,** 075202 (2011).
47) F. Bruneval and G. Roma, Physical Review B **83,** 144116 (2011).
48) M. Bockstedte, A. Gali, A. Mattausch, O. Pankratov, and J. W. Steeds, Physica Status Solidi B-Basic Solid State Physics **245,** 1281 (2008).
49) H. Itoh, A. Kawasuso, T. Ohshima, M. Yoshikawa, I. Nashiyama, S. Tanigawa, S. Misawa, H. Okumura, and S. Yoshida, Physica Status Solidi a-Applied Research **162,** 173 (1997).
50) T. Liao, G. Roma, and J. Y. Wang, Philosophical Magazine **89,** 2271 (2009).
51) E. Rauls, T. Frauenheim, A. Gali, and P. Deak, Physical Review B **68,** 155208 (2003).
52) G. Roma and J. P. Crocombette, Journal of Nuclear Materials **403,** 32 (2010).
53) U. Gerstmann, A. P. Seitsonen, D. Ceresoli, F. Mauri, H. J. von Bardeleben, J. L. Cantin, and J. G. Lopez, Physical Review B **81,** 195208 (2010).
54) S. T. Nakagawa, A. Okamoto, and G. Betz, Nuclear Instruments & Methods in Physics Research Section B-Beam Interactions with Materials and Atoms **266,** 2711 (2008).
55) T. A. G. Eberlein, C. J. Fall, R. Jones, P. R. Briddon, and S. Oberg, Physical Review B **65,** 184108 (2002).
56) M. Posselt, F. Gao, and W. J. Weber, Physical Review B **73,** 125206 (2006).
57) F. Gao, J. Du, E. J. Bylaska, M. Posselt, and W. J. Weber, Applied Physics Letters **90,** 221915(2007).
58) T. A. G. Eberlein, R. Jones, S. Oberg, and P. R. Briddon, Physical Review B **74,** 144106 (2006).
59) T. Oda, Y. W. Zhang, and W. J. Weber, Journal of Chemical Physics **139,** 124707 (2013).
60) F. Gao, W. J. Weber, H. Y. Xiao, and X. T. Zu, Nuclear Instruments & Methods in Physics Research Section B-Beam Interactions with Materials and Atoms **267,** 2995 (2009).
61) A. Gali, P. Deak, P. Ordejon, N. T. Son, E. Janzen, and W. J. Choyke, Physical Review B **68,** 125201 (2003).
62) A. Mattausch, M. Bockstedte, and O. Pankratov, Physical Review B **70,** 235211 (2004).
63) C. Jiang, D. Morgan, and I. Szlufarska, Acta Materialia **62,** 162 (2014).
64) H. Jiang, C. Jiang, D. Morgan, and I. Szlufarska, Computational Materials Science **89,** 182 (2014).
65) T. Hornos, N. T. Son, E. Janzen, and A. Gali, Physical Review B **76,** 165209 (2007).
66) T. Liao and G. Roma, Nuclear Instruments & Methods in Physics Research Section B-Beam Interactions with Materials and Atoms **327,** 52 (2014).
67) F. Schmid, S. A. Reshanov, H. B. Weber, G. Pensl, M. Bockstedte, A. Mattausch, O. Pankratov, T. Ohshima, and H. Itoh, Physical Review B **74** (2006).
68) W. J. Weber, Nuclear Instruments & Methods in Physics Research Section B-Beam Interactions with Materials and Atoms **166,** 98 (2000).
69) N. Swaminathan, P. J. Kamenski, D. Morgan, and I. Szlufarska, Acta Materialia **58,** 2843 (2010).
70) Y. Zhang, F. Gao, W. Jiang, D. E. McCready, and W. J. Weber, Physical Review B **70,** 125203 (2004).
71) V. Nagesh, J. W. Farmer, R. F. Davis, and H. S. Kong, Applied Physics Letters **50,** 1138 (1987).
72) L. Patrick and W. J. Choyke, Journal of Physics and Chemistry of Solids **34,** 565 (1973).
73) J. A. Freitas, S. G. Bishop, J. A. Edmond, J. Ryu, and R. F. Davis, Journal of Applied Physics **61,** 2011 (1987).
74) J. Rottler, D. J. Srolovitz, and R. Car, Physical Review B **71,** 064109 (2005).
75) D. J. Bacon and T. D. Delarubia, Journal of Nuclear Materials **216,** 275 (1994).